\documentclass[preprint,showpacs,aps,preprintnumbers,amsmath,amssymb,groupedaddress]{revtex4}

\usepackage[dvips]{graphicx}
\usepackage{dcolumn}
\usepackage{bm}
\usepackage[latin1]{inputenc}

\begin{document}

%
%
\title{Solution of the inverse problem in spherical gravitational wave detectors using a model with independent bars}

\email{chlenzi@ita.br}
\email{nadjam@cefetsp.br}
\email{marinho@ita.br}

\author{C\'{e}sar H. Lenzi$^{1,2*}$, Nadja S. Magalh\~aes$^2$, Rubens M. Marinho Jr.$^{1\dagger}$, C\'{e}sar A. Costa, Helmo A. B. Ara\'{u}jo$^1$ and Odylio D. Aguiar$^{3}$}
\affiliation{$^1$Departamento de F\'{i}sica, Instituto Tecnol\'{o}gico de  Aeron\'{a}utica,
Campo  Montenegro, S\~{a}o Jos\'{e} dos Campos,
SP, 12228-900, Brazil \\
$^2$Departamento de F\'{i}sica, Universidade de Coimbra, Rua Larga, Coimbra, 3004-516, Portugal \\
$^3$Centro Federal de Educa\c{c}\~ao de Tecnol\'ogica de S\~ao Paulo, R. Dr. Pedro Vicente 625, S\~ao Paulo, SP 01109-010, Brazil \\
$^4$Departamento de Astrof\'{i}sica, Instituto Nacional de Pesquisas Espaciais,  Av. dos Astronautas 1.758,  S\~{a}o Jos\'{e} dos Campos, SP, 12227-010, Brazil}
\date{\today}

\begin{abstract}

The direct detection of gravitational waves will provide valuable
astrophysical information about many celestial objects. The SCHENBERG
has already undergone its first test run. It is expected
to have its first scientific run soon. In this work  a new data
analysis approach is presented, called method of independent bars,
which can be used with SCHENBERG's data . We test this method
through the simulation of the detection of gravitational waves.
With this method we find the source's direction without the need
to have all six transducers operational. Also we show that the
method is a generalization of another one, already described in
the literature, known as the mode channels method.
\end{abstract}


\maketitle

\section{\label{sec1}Introduction\protect}

The detection of gravitational waves will have many implications
in Physics and Astrophysics. Besides the confirmation of the
general relativity theory, it will allow the investigation of
several astrophysical phenomena, such as the existence of black
holes and the mass and abundance of neutron stars, thus allowing
new scientific frontiers.

Joseph Weber was the first one to propose feasible gravitational
wave detectors in the 1960's\cite{weber}. The antenna he proposed
had a cylindrical shape, operated at room temperature and was
acoustically isolated. Since his pioneering work the detectors
based on resonant-mass antennas have improved significantly. For
the latest generation of such detectors, the antenna has a spherical
shape.

The transducer distribution on spherical detectors follows the
truncated icosahedron configuration, proposed by Johnson and
Merkowitz in 1993\cite{merk93} and justified formally by
Magalh\~aes et al. in 1997\cite{maga97b}. Spherical antennas have
omnidirectional sensitivity and all observables needed for
gravitational astronomy can be obtained from only one spherical
detector appropriately equipped with six transducers\cite{maga95}.
When fully operational, such detector will be able not only to
acknowledge the presence of a gravitational wave within its
bandwidth: it will be able to inform the direction of a source in
the sky as well as the amplitudes of the wave's polarization
components. This is the case of SCHENBERG, the second spherical
detector ever built in the world and the first equipped with a set
of parametric transducers, which is installed at the Physics
Institute of the University of Sao Paulo (at Sao Paulo city,
Brazil). It has undergone its first test run in September 8, 2006,
with three transducers operational. Recent
information on the present status of this detector can be found in
reference \cite{cqg2008}.

There are mathematical models for this detector for the case that
all six transducers are operational. Such models investigated two
situations: one in which the transducers are perfectly
uncoupled\cite{maga97b,merk97,cesar06} and another in which the
transducers are somehow coupled to each other\cite{merk99}. In both cases it is possible to solve the
inverse problem and retrieve the four astrophysical parameters of
interest: the angles that determine the source's position in the
sky, denoted by $\beta$ (zenith angle) and $\gamma$ (azimuthal
angle), and the amplitudes of the two polarization states as
functions of time, $h_+(t)$ and $h_\times(t)$.

In this work we present a methodology for the solution of the
inverse problem in spherical detectors, which we named ``the
method of independent bars". Unlike the previous methods, it
allows the determination of astrophysical parameters using less
than six transducers. In fact, we will show that it generalizes a
previous approach, known as the mode channels method. We tested
our method with simulations and found encouraging results. In the
next sections we will show these results in detail.

\section{The Method of Independent Bars}

The determination of directions is sky demands the use of
appropriate reference frames. For this reason we shall define two
frames: the first one is the {\it laboratory system} with axes
defined by the unit vectors $({\mathbf e}_x, {\mathbf e}_y,
{\mathbf e}_z)$. The second frame is the {\it gravitational wave
system}, which will have its axes defined by $({\mathbf e}_{X},
{\mathbf e}_{Y}, {\mathbf e}_{Z})$ with the wave travelling in the
${\mathbf e}_{Z}$ direction and the axes of the polarization
amplitudes being $({\mathbf e}_{X}, {\mathbf e}_{Y})$.

The projection of the polarization state amplitudes of the
gravitational signal on the shaft of a detector can be expressed
by the contraction of two symmetric, trace-free
tensors\cite{maga97b}:
\begin{equation}\label{eq1}
R(t) =\frac{1}{2} h^{ij}(t) D_{ij}.
\end{equation}
The response amplitude of one transducer
coupled to the spherical detector is given by $R(t)$, while $
h^{ij}(t)$ is the spatial part of the gravitational wave tensor
and $D_{ij}$ is the transducer's tensor, which informs the
orientation of the transducer on the spherical detector.

In Appendix \ref{appen2} we show that the contraction in equation
(\ref{eq1}) is a representation of the time-dependent scalar
potential given by\cite{merk97}:
\begin{equation}\label{potential}
\phi({\bf x},t)=\frac{1}{2}\rho x_j \ddot{h}^{jk}(t)x_k,
\end{equation}
where $\rho$ is the antenna's mass density and $x_i$ is a
coordinate location.

\subsection{The Transducer's Tensor}

It is possible to model the system as if we had one independent
bar behind each transducer on the detector\cite{maga95}. Such a
bar detector, whose axis is in the ${\mathbf n}$ direction, can be
characterized by the symmetric tensor
\begin{equation}\label{eq2}
D_{ij} = ({\mathbf n}\otimes {\mathbf n})_{ij}.
\end{equation}

The transducers are located on the antenna and the position of the
$L^{th}$ transducer is described by the unit vector
${\mathbf{n}}$.  In the laboratory system (lab frame) this vector
is given by
\begin{equation}
{\mathbf n}_L = \sin \theta_L \cos \phi_L\mathbf e_x+ \sin \theta_L\sin \phi_L\mathbf e_y+ \cos \theta_L\mathbf e_z,
\label{n}
\end{equation}
where $\theta_L$ and $\phi_L$ are angles that locate the
transducer on the spherical detector. Using equations (\ref{eq2}) and
(\ref{n}) the $L^{th}$ transducer's tensor assume the following
form in the lab frame:
\begin{equation}\label{eq3}
\mathbf D^L = \begin{bmatrix}
\sin^2\theta_L\cos^2\phi_L  & \frac{1}{2}\sin^2\theta_L\sin2\phi_L    & \frac{1}{2}\sin2\theta_L\cos\phi_L                                                  \cr
\frac{1}{2}\sin^2\theta_L\sin2\phi_L &  \sin^2\theta_L\sin^2\phi_L            &\frac{1}{2}\sin2\theta_L\sin\phi_L                                                \cr
\frac{1}{2}\sin2\theta_L\cos\phi_L  & \frac{1}{2}\sin2\theta_L\sin\phi_L     &\cos^2\theta_L \cr
\end{bmatrix}.
\end{equation}

\subsection{The Wave's Tensor}

Consider the complex null vector $\mathbf m$ defined by
\cite{Newman}:
\begin{equation}
{\mathbf m} = \frac{1}{\sqrt{2}}\left({\mathbf e}_{X} + i {\mathbf e}_{Y}\right),
\end{equation}
where ${\mathbf e}_{X}$ and ${\mathbf e}_{Y}$ are unit vectors in
the wave frame. The gravitational wave tensor $\mathbf h$ can be
written in terms of  these vectors:
\begin{equation}\label{}
\mathbf h = 2h_+\Re({\mathbf m}\otimes{\mathbf m}) +
2h_\times\Im({\mathbf m}\otimes{\mathbf m})
\end{equation}
or
\begin{equation}\label{eqwave2}
\mathbf h = h_+\mathbf e_+ +h_\times\mathbf e_\times,
\end{equation}
where
\begin{align*}
        \mathbf{e}_+ &=\mathbf{e}_{X}\otimes\mathbf{e}_{X}-\mathbf{e}_{Y}\otimes\mathbf{e}_{Y},\\
        \mathbf{e}_\times &=\mathbf{e}_{X}\otimes\mathbf{e}_{Y}+\mathbf{e}_{Y}\otimes\mathbf{e}_{X}.
\end{align*}

Starting from the lab frame $(\mathbf e_x,\mathbf e_y, \mathbf
e_z)$ it is possible to rotate by $\gamma$ around $\mathbf e_z$ to
obtain  $(\mathbf e_{x'},\mathbf e_{y'}, \mathbf e_{z'})$ and
rotate by $\beta$ around  $\mathbf e_{y'}$ (Euler $y$-convention)
to obtain $(\mathbf e_{x''},\mathbf e_{y''}, \mathbf e_{z''})$ so
that  $\mathbf e_{z''}$ points in the direction of the propagation
of the wave $\mathbf e_{Z}$.

 In the lab frame the vector $\mathbf m$ thus takes the
following form:
\begin{equation}
{\mathbf m} = \frac{1}{\sqrt{2}}
\left[
        (\cos\beta\cos\gamma-i\sin\gamma){\mathbf e}_x +
        (\cos\beta\sin\gamma+i\cos\gamma){\mathbf e}_y -
        \sin\beta {\mathbf e}_z
\right],
\end{equation}
where $\beta$, $\gamma$ are the zenithal and azimuthal angles,
respectively, that define the direction of arrival of the
gravitational signal.

We have determined the expressions for the wave's and the
detector's tensors relative to the laboratory system. We can now
solve the inverse problem.

\subsection{The inverse problem}
Equation (\ref{eq1}), valid for one transducer, can be expanded as \cite{cesar0801}
\begin{equation}\label{eq4}
\frac{1}{2}\left[D_{11}^Lh_{11} + 2D_{12}^Lh_{12} + 2D_{13}^Lh_{13}
+ D_{22}^Lh_{22} + 2D_{23}^Lh_{23} + 2D_{33}^Lh_{33}\right] = R^L.
\end{equation}

In the case that we have six operating transducers, using the six
independent components of the spatial part of the gravitational
waves tensor, the $h_{ij}$ can be obtained as the solutions of the
system of equations
\begin{equation}
\begin{bmatrix}
D_{11}^1 & 2D_{12}^1 & 2D_{13}^1 & D_{22}^1 & 2D_{23}^1 & D_{33}^1 \cr
D_{11}^2 & 2D_{12}^2 & 2D_{13}^2 & D_{22}^2 & 2D_{23}^2 & D_{33}^2 \cr
D_{11}^3 & 2D_{12}^3 & 2D_{12}^3 & D_{22}^3 & 2D_{23}^3 & D_{33}^3 \cr
D_{11}^4 & 2D_{12}^4 & 2D_{13}^4 & D_{22}^4 & 2D_{23}^4 & D_{33}^4 \cr
D_{11}^5 & 2D_{12}^5 & 2D_{13}^5 & D_{22}^5 & 2D_{23}^5 & D_{33}^5 \cr
D_{11}^6 & 2D_{12}^6 & 2D_{13}^6 & D_{22}^6 & 2D_{23}^6 & D_{33}^6
\end{bmatrix}
\left[
\begin{array}{c}
h_{11} \cr
h_{12} \cr
h_{13} \cr
h_{22} \cr
h_{23} \cr
h_{33} \cr
\end{array}
\right] =
2\left[
\begin{array}{c}
R^1 \cr
R^2 \cr
R^3 \cr
R^4 \cr
R^5 \cr
R^6
\end{array}
\right].
\label{sixt}
\end{equation}
This system allow us to calculate the components of the wave's tensor 
and to check the transversality and tracelessness of the wave tensor,
which will be present if GR is the correct theory of gravitation. 
On the other hand, in the case the detector is coupled to a set of
only five transducers ($ L = 1,\ldots 5$), the last line will be
absent. So in its place we use the trace-free condition
\cite{cesar0801}:
\begin{equation}
h_{11} + h_{22} + h_{33} = 0.
\label{tracefree}
\end{equation}

When only four transducers are operational it is still possible
to solve the system of equations by adding to it the
transversality condition of the gravitational wave. This condition
implies that the spatial part of the gravitational wave tensor is
orthogonal to the direction of propagation, i.e., $h_{ij}n^j=0$.
In other words, the matrix $h_{ij}$ has one eigenvalue equal to
zero and, as a consequence, null determinant. We then use the
condition of null determinant, given by
\begin{equation}
h_{11}h_{22}h_{33}+2h_{12}h_{13}h_{23}-h_{33}h_{12}^2-h_{22}h_{13}^2-h_{11}h_{23}^2=0,
\label{transv}
\end{equation} instead of the line that is
missing in the system of equations (\ref{sixt}).

Since we take General Relativity for granted the $h$
matrix, without noise, must be ``TT": transverse, equation (\ref{transv}), and trace free, equation (\ref{tracefree}). In this case
any set of 6 $h_{\mu \nu}$, obtained as described above,  should
obey such condition. The presence of noise is expected to make the
$h_{\mu \nu}$ vary so that the transversality and tracelessness of
$\mathbf h$ is not perfect. Of course, if the noise is too high it will
be difficult to identify the ``TT'' properties in the $\mathbf h$ matrix. Our
ability to extract the signal from the noise will depend on the
signal-to-noise ratio achieved for the particular event. This has
been investigated by Merkowitz \cite{Merkowitz1998}
and Merkowitz, Lobo and Serrano \cite{Merkowitz1999}, partly using an approach
first presented in Magalhaes et al. \cite{Nadja1997}. In the present work we will not
approach this issue in depth, which will be left to be discussed
elsewhere.

\section{The astrophysical parameters}
With the $\mathbf h $ obtained from the the system of equations
(\ref{sixt}) is possible to calculate all the relevant parameters
of the gravitational wave: the angles $\beta$ and $\gamma$ that
indicate the direction of the wave and the amplitudes of the two
polarization states of the wave as functions of time $h_+(t)$ and
$h_\times(t)$.

\subsection{The direction of the gravitational wave}
For a gravitational wave that propagates in the direction of the
${\mathbf e}_Z$ axis of its proper system of coordinates, in
the case of general relativity the transversality condition
relation holds: \cite{Tinto,maga97c}
\begin{equation}\label{eq7}
\mathbf h\centerdot{\mathbf e}_{Z} = 0.
\end{equation}

The  wave frame can be written in terms of the laboratory frame
through a rotation. Following the $y$-convention for the Euler
angles, $\mathbf e_{Z}$ assumes the form:
\begin{equation}\label{eq8}
{\mathbf e}_{Z}= \sin\beta\cos\gamma\,\mathbf e_x+ \sin\beta\sin\gamma\,\mathbf e_y+ \cos \beta\,\mathbf e_z.
\end{equation}
Substituting equation (\ref{eq8}) in (\ref{eq7}) and using the
symmetry of $h_{ij}$ we obtain the following system of equations
\begin{subequations}
\begin{align}
h_{11}\sin\beta\cos\gamma + h_{12}\sin\beta\sin\gamma + h_{13}\cos\gamma = 0,
\label{cond1}
\\
h_{12}\sin\beta\cos\gamma + h_{22}\sin\beta\sin\gamma + h_{23}\cos\gamma = 0,
\label{cond2}
\\
h_{13}\sin\beta\cos\gamma + h_{23}\sin\beta\sin\gamma + h_{33}\cos\gamma = 0.
\label{cond3}
\end{align}
\end{subequations}
Since $\det \mathbf h=0$, in this system the equations are
linearly dependent. Two of them are used to obtain $\beta$ and
$\gamma$:
\begin{equation}\label{dirgamma}
 \gamma = \frac{h_{12}h_{13} - h_{23}h_{11}}{h_{12}h_{23} - h_{13}h_{22}},
\end{equation}

\begin{equation}\label{dirbeta}
\tan \beta = \pm\frac{h_{12}h_{23} - h_{13}h_{22}}{h_{11}h_{22} - h_{12}^2}\sqrt{1+\tan^2\gamma}.
\end{equation}
These two solutions for $\beta$ are related to the two
diametrically opposed positions in the sky from which the
gravitational wave can arrive.

\begin{figure}[!t]
\begin{center}
\includegraphics[width = 15.5cm,height=15.5cm]{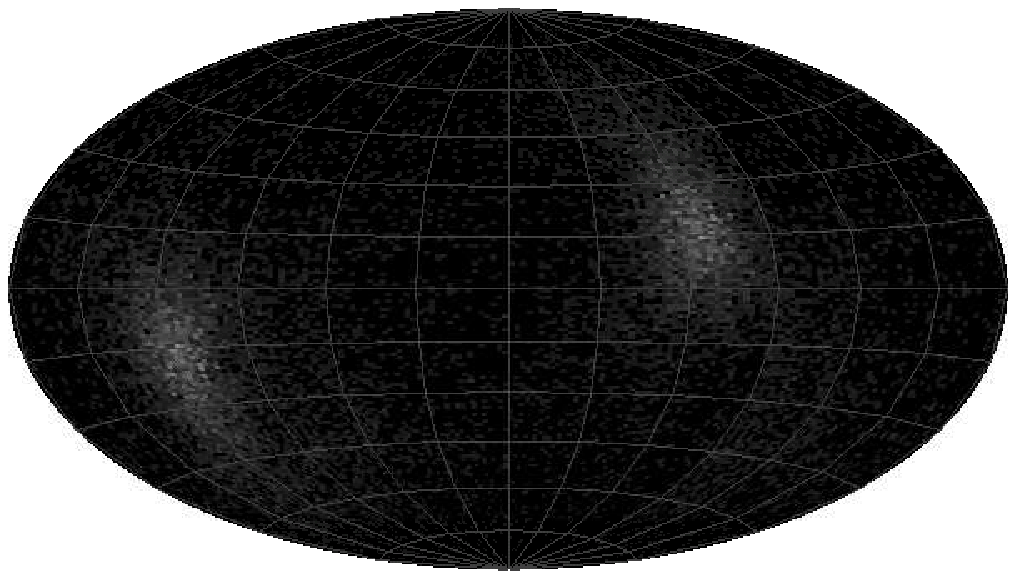}
\caption{Determination of the wave's direction using the method of independent bars the presence of Gaussian noise with $SNR\sim8$. The transducers used are numbers 1, 2, 3, 4 and 6 in Table \ref{angulosit}).}
\label{result2}
\end{center}
\end{figure}

\begin{figure}[!t]
\includegraphics[width = 15.5cm,height=15.5cm]{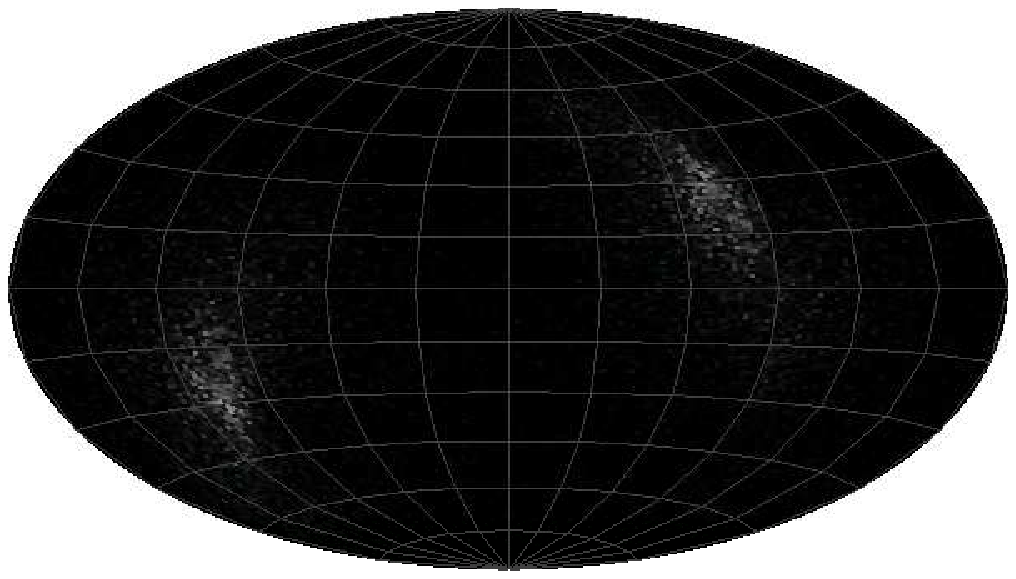}
\caption{Determination of the wave's direction uning the method of independent bars in the presence of Gaussian noise with $SNR\sim8$. The six transducers are positioned according to the truncated icosahedro configuration shown in Table \ref{angulosit}.}
\label{result3}
\end{figure}

With the above equations we compute the
direction of the incoming wave. It results in the same value for
each time sample, in the case of a pure (noiseless) signal. In the
case of a noisy signal, a distribution around the mean values of
the $\beta$ and $\gamma$ angles is found, as we can see from
Figures \ref{result2} and \ref{result3}.

\subsection{Amplitudes of the polarization states}
Using the fact that $\mathbf e_+\centerdot\mathbf e_+=2$, $\mathbf
e_\times\centerdot\mathbf e_\times=2$ and $\mathbf
e_+\centerdot\mathbf e_\times=0$ we obtain $h_+$ and $h_\times$
with the aid of equation (\ref{eqwave2}), yielding
\begin{align*}
        h_+ = &(\cos^2\beta\cos^2\gamma -\sin^2\gamma) h_{11} + \sin2\gamma(1+\cos^2\beta) h_{12} - \sin2\beta\cos\gamma h_{13}+ \\
              &(\cos^2\beta\sin^2\gamma -\cos^2\gamma) h_{22} - \sin2\beta\sin\gamma h_{23}  + \sin^2\beta h_{33},\\
        h_\times= & -\cos\beta\sin2\gamma h_{11}+2\cos\beta\cos2\gamma h_{12}+2\sin\beta\sin\gamma h_{13}\\
                &+\cos\beta\sin2\gamma h_{22}  -2\sin\beta\cos\gamma
                h_{23}.
\end{align*}

Once these amplitudes are known other quantities can be
calculated. One of them is the polarization angle, $\Psi$, through
the relation
\[
\tan(2\psi(t)) = \frac{h_x(t)}{h_+(t)}.
\]
However, this parameter is actually not relevant for gravitational
wave detection purposes since it is just an arbitrary angle
between the wave x axis and an axis where the wave has only the
instantaneous plus polarization, $h_+$.

On the other hand, important information can be obtained from the
difference between the polarization phases. This can be determined
if we analyze the time evolution of the instantaneous
polarizations registered in each sample (the polarizations are
time dependent). Our ability to determine this phase difference
depends on the signal to noise ratio for each particular event
sampled. Such time varying analysis is beyond the scope of the
present work.

\section{Testing the method of independent bars}
\begin{figure}[ht]
\begin{center}
\includegraphics[width = 10.5cm]{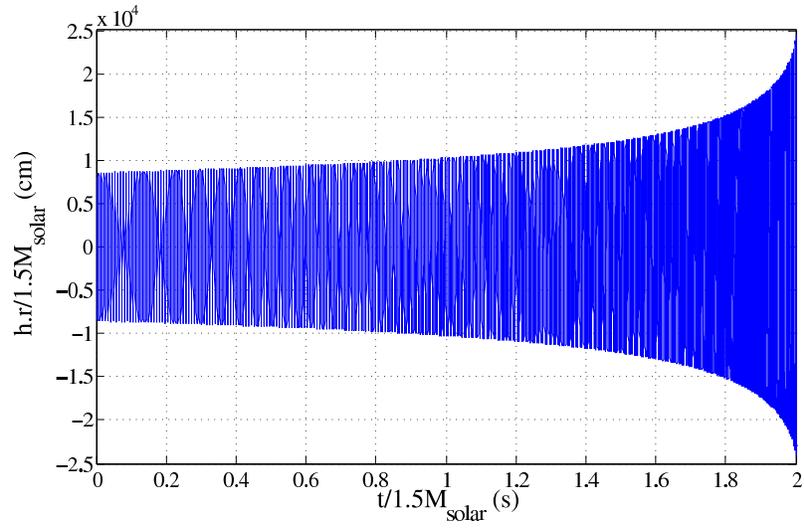}
\caption{Waveform of a gravitational wave emitted by a binary
neutron star-neutron star system near its last stable orbits. The
time and amplitude axes are scaled according to the object masses
through the multiplication by $M/1.5M_{solar}$, where $M_{solar}$
is the Sun's mass.} \label{Duez}
\end{center}
\end{figure}

With the objective of testing the efficiency of the method we
produced simulated data of the SCHENBERG detector. For the
waveform we chose a template from the Laboratory for High Energy
Astrophysics (LHEA) of the NASA/GFSC\cite{astrogravs},
one of the three waveforms used by Duez and collaborators
\cite{duez01,duez01a} and depicted in Figure \ref{Duez}.

This signal was introduced in a simulator of
SCHENBERG\cite{cesartese}, that returns the response of the six
two-mode transducers positioned on the detector their positions
obeying the truncated icosahedral configuration\cite{ces04}.
The simulation of the detector is possible because its
mathematical model is known as well as the mathematical models of
its noise sources \cite{Frajuca2004}.

Table \ref{angulosit} shows the angles relative to the position of
each transducer on the spherical detector. We simulated the
transducers' responses when detector was excited by a wave of the
kind shown in Figure \ref{Duez} arriving at the detector in the
direction defined by $\beta = 30^\circ$ and $\gamma = 75^\circ$.

\begin{table}[ht]
\begin{center}
\caption{The angle $\theta_L$ is the one that a line L, that passes
through the center of a pentagonal face and the origin of the lab
frame, makes with the local zenith and $\phi_L$ is the angle that L
makes with the $x$ axis of the lab frame.}
\begin{tabular}{c c c c c c c c}  \hline \hline
  $L$    & \ \ \ \ \ \ &    1          &     2    &    3     &   4      &    5     &    6     \\  \hline
$\theta_L$ & \ \ \ \ \ \ & $79,1877^\circ$ & $79,1877^\circ$ & $79,1877^\circ$ & $37,3774^\circ$ & $37,3774^\circ$ & $37,3774^\circ$  \\ \hline
$\phi_L$     & \ \ \ \ \ \ &   $120^\circ$     &      $240^\circ$   &       $0^\circ$      &     $300^\circ$    &     $180^\circ$    &   $60^\circ$    \\ \hline \hline
\label{angulosit}
\end{tabular}
\end{center}
\end{table}




In the SCHENBERG simulator the amplitude signal-to-noise ratio (SNR) 
was estimated from Thorne as \cite{thornebook}
\begin{equation}
SNR=\sqrt{\int_{-\infty}^{\infty}\tilde{W}(f)\frac{|\tilde{h}^{GW}(f)|^2}{S_N(f)}df},
\end{equation}
where $\tilde{W}(f)$ represents the applied filter function
(chosen, in the present work, as the function of a pass-band
filter). $S_N(f)$ is the equivalent noise profile (Gaussian, in
this case); it 
can be obtained from the detector's output when no useful signal
is present, representing the total noise energy in the sphere.

The signal-to-noise 
used was $SNR \sim 8$ 
corresponding to a source %
with behavior as in Figure \ref{Duez}, 
located approximately 
$97$ kpc away. It is from this source model that
$\tilde{h}^{GW}(f)$ can be obtained.


We used the output of 5 
or 6 transducers to test our model, as follows.

\subsection{Case 1: pure signal}

Once we had the simulated response of the transducers we
calculated the parameters of the gravitational wave using the
method of independent bars.

The result obtained with the method  for the source's direction
was exactly the same initially imposed: $\beta = 30^\circ$ and
$\gamma = 75^\circ$. This shows that the method has good accuracy
and does not present intrinsic errors.

\subsection{Case 2: signal plus noise}

We then tested the method
in the
presence of noise 
adding Gaussian noise to the data. To
illustrate the results we projected the sphere 
onto a plane using 
the Hammer-Aitoff projection\cite{Calabretta}, where the direction
is given by $\theta=90^\circ-\beta$ and $\phi=-\gamma$. 

Figures \ref{result2} and \ref{result3} show the source's
direction as found using the method of independent bars for the
cases 
where the data of six 
or five transducers are used. 
In those 
figures the two regions of the sky are marked because there is a
natural ambiguity in the determination of the direction by only
one detector. If more than
one detector is used then 
there is a time delay between detections and one can determine
whether the wave came from up or down.





As one can verify, the smaller the the number of 
transducers' data used in the analysis the smaller the precision 
of the results. In 
those figures we observe that, for the case 
in which the data of only five transducers are used, the area of
interest is larger than in the case 
that use the data of all six transducers.
This area is equivalent
to the error. The numerical results obtained for the method are
presented in Table \ref{results}.
\begin{table}[ht]
\caption{Numerical results obtained with the method of independent
bars for the wave's direction in the case of signal in the
presence of noise ($SNR \sim 8$) using five 
or six
transducers' data. $n$ is the number of the transducers.}
\begin{center}
\begin{tabular}{c c c}  \hline \hline
 $n$ & $\beta$ & $\gamma$ \\ \hline
 \ \ \ \ 5 \ \ \ \ & \ \ \ \ $(27.2\pm 7.1)^\circ$ \ \ \ \ & \ \ \ \ $(78.7\pm 5.3)^\circ$ \ \ \ \ \\ \hline
 \ \ \ \ 6 \ \ \ \ & \ \ \ \ $(28.1\pm 4.2)^\circ$ \ \ \ \ & \ \ \ \ $(75.3\pm 2.9)^\circ$ \ \ \ \ \\ \hline
\end{tabular}
\end{center}
 \label{results}
\end{table}

The values presented in this Table were obtained simulating the
wave's incidence 50 times and then calculating the average and
the standard deviation of the results.

We have verified the efficiency of the method of independent bars
in determining a source's direction. Another important result is
to show the equivalence between the method of  independent bars
and mode channels model \cite{merk97}. This will be done in the
next section.

\section{Equivalence between the method of independent bars and the
mode channels model}

In order to show the equivalence between the mathod of independent bars and mode channels model, we must write the transducer's tensor (eqs. \ref{eq2} and \ref{eq3}) in its trace-free form:
\begin{equation}\nonumber
D_{ij} = ({\mathbf n}\otimes {\mathbf n})_{ij} - \frac{1}{3}\delta_{ij},
\end{equation}
where the term $\frac{1}{3}\delta_{ij}$ ensures the property of
tracelessness of the tensor.

Each component of the matrix ${\mathbf D}$ (eq. \ref{sixt}) is a function of the angles
$\theta_L$ and $\phi_L$ that informs the position of each
transducer. When we impose the trace-free condition on the tensor $D_{ij}$, all
this functions can be written as a combination of real
spherical harmonics of second order. Therefore the matricial system
can be written as
\begin{equation}\label{solproinv3}
\left[
\begin{array}{c}
R^1 \cr
R^2 \cr
R^3 \cr
R^4 \cr
R^5 \cr
R^6
\end{array}
\right] =\frac{1}{2}\sqrt{\frac{16 \pi}{15}}
\begin{bmatrix}
Y_{1}^1 & Y_{2}^1 & Y_{3}^1 & Y_{4}^1 & Y_{5}^1 \cr
Y_{1}^2 & Y_{2}^2 & Y_{3}^2 & Y_{4}^2 & Y_{5}^2 \cr
Y_{1}^3 & Y_{2}^3 & Y_{2}^3 & Y_{4}^3 & Y_{5}^3 \cr
Y_{1}^4 & Y_{2}^4 & Y_{3}^4 & Y_{4}^4 & Y_{5}^4 \cr
Y_{1}^5 & Y_{2}^5 & Y_{3}^5 & Y_{4}^5 & Y_{5}^5 \cr
Y_{1}^6 & Y_{2}^6 & Y_{3}^6 & Y_{4}^6 & Y_{5}^6 \cr
\end{bmatrix}
\left[
\begin{array}{c}
h_{1}\cr h_{2} \cr h_{3} \cr h_{4} \cr h_{5} \cr
\end{array}
\right],
\end{equation}
where $h_m$ are the quadrupolar representation of the
gravitational waves. In Appendix \ref{appen1} we shown the details
of this calculation.

When the transducer are positioned according to the truncated
icosahedral configuration the matrix of the real spherical
harmonics $\mathbf Y$ is the transpose of the model matrix
${\mathbf B}^T_{6\times5}$ proposed by Merkowitz and
Johnson\cite{Johnson}.  In compact notation equation
(\ref{solproinv3}) can be written as
\begin{equation}
{\mathbf R} = \frac{1}{2}\sqrt{\frac{16\pi}{15}}{\mathbf
B}^T_{6\times5} {\mathbf h}.
\end{equation}
Using a property of the model
matrix\cite{merk97}:
\begin{equation}
{\mathbf B}_{5\times6}{\mathbf B}^T_{6\times5} = \frac{3}{2\pi} {\mathbf I},
\end{equation}
we obtain:
\begin{equation}
{\mathbf h}= \sqrt{\frac{5 \pi}{3}} {\mathbf B}_{5\times6} {\mathbf
R},
\end{equation}
or,  in explicit form:
\begin{equation}\label{solproinv4}
\left[
\begin{array}{c}
h_{1} \cr h_{2} \cr h_{3} \cr h_{4} \cr h_{5} \cr
\end{array}
\right] = \sqrt{\frac{5\pi}{3}}
\begin{bmatrix}
Y_{1}^1 & Y_{1}^2 & Y_{1}^3 & Y_{1}^4 & Y_{1}^5 & Y_{1}^6 \cr
Y_{2}^1 & Y_{2}^2 & Y_{2}^3 & Y_{2}^4 & Y_{2}^5 & Y_{2}^6 \cr
Y_{3}^1 & Y_{3}^2 & Y_{3}^3 & Y_{3}^4 & Y_{3}^5 & Y_{3}^6 \cr
Y_{4}^1 & Y_{4}^2 & Y_{4}^3 & Y_{4}^4 & Y_{4}^5 & Y_{4}^6 \cr
Y_{5}^1 & Y_{5}^2 & Y_{5}^3 & Y_{5}^4 & Y_{5}^5 & Y_{5}^6 \cr
\end{bmatrix}
\left[
\begin{array}{c}
R^1 \cr
R^2 \cr
R^3 \cr
R^4 \cr
R^5 \cr
R^6
\end{array}
\right].
\end{equation}

This result, obtained with the independent bars method, coincides
with the one found with the mode channels model. Therefore, if the
transducers are in a truncated icosahedron configuration the two
models are equivalent.

On the other hand, it is worth stressing that even if the
transducers are not in this special configuration the inverse
problem can be solved using the independent bars model.

\section{Summary}

The main objective of this work was to show the details of the
analytical construction of the method of independent bars and its
efficiency in the determination of the
relevant parameters of gravitational waves detected by spherical
detector, namely: the two angles that determine the source's
position in the sky  and the amplitudes of the two polarization
states as functions of time.

As can be seen from the results, the method of independent bars is
efficient for the solution of the inverse problem. First we tested
the method in the noiseless case  and we verified that it does not
present  intrinsic errors, being thus an exact method. Then we
tested the method adding Gaussian noise to the data, as can be
seen in Figures \ref{result2} and \ref{result3}.

A major advantage of the method of independent bars is its
generality, since it does not depend on the specific transducers'
configuration. The fact that $\mathbf D$ is a square matrix
facilitates the inversion of the system's equation, the only
requirement being that the matrix is non-singular.

Therefore, the usefulness of this method is evident. For instance,
in the absence of any transducer coupled to the spherical detector
the mode channels model\cite{merk97} does not work. This happens
because the truncated icosahedron configuration is lost. However,
the independent bars method still works under this condition. We
are currently investigating ways to solve the inverse problem with
three transducers. We have already got some preliminary results
that will appear in a future paper.

In this work we have actually shown that the method of the
independent bars is a generalization of the mode channels model.
The matrix $\mathbf D$ can be written in terms of the real
spherical harmonics. In this case, if the transducers' positions
obey the truncated icosahedron configuration we verified that
$\mathbf D$ is the model matrix $\mathbf{B}_{5\times6}$ multiplied
by a constant.

Finally, we point out that this method can be used for the
solution of the inverse problem in a network of bar antennas, as
long as a common reference frame is chosen and time delays are
accounted for.

\begin{acknowledgements}
The authors thank the financial support given by their respective
Brazilian funding agencies: CHL to CAPES by the fellowship 2071/07-0 and the international cooperation program Capes-Grices between Brazil-Portugal. NSM and RMMJ to FAPESP (grants \# 2006/07316-0 and
\# 07/51783-4). A special acknowledgement is given to FAPESP for
supporting the construction and operation of the SCHENBERG
detector (grant \# 2006/56041-3, Thematic Project: ``New Physics
from Space: Gravitational Waves'').
\end{acknowledgements}

\newpage 

\clearpage

\appendix

\section{}\label{appen1}

Each component of the matrix (\ref{sixt}) is a function of the
angles $\theta$ and $\phi$ that inform the position of each
transducer. The $D$ functions can be written as combinations of
the real spherical harmonics as follows:
\begin{subequations}
\begin{align}
D_{11}^L &
=\frac{1}{2}\left(\sqrt{\frac{16\pi}{15}}Y_1^L-\sqrt{\frac{16\pi}{45}}Y_5^L
\right), \nonumber
\\\nonumber
D_{12}^L & =-\frac{1}{2}\sqrt{\frac{16\pi}{15}}Y_2^L,
\\\nonumber
D_{13}^L & =\frac{1}{2}\sqrt{\frac{16\pi}{15}}Y_3^L,
 \\\nonumber
D_{22}^L &
=-\frac{1}{2}\left(\sqrt{\frac{16\pi}{15}}Y_1^L+\sqrt{\frac{16\pi}{45}}Y_5^L
\right),
\\\nonumber
D_{23}^L & =-\frac{1}{2}\sqrt{\frac{16\pi}{15}}Y_4^L,
\\\nonumber
D_{33}^L & =\sqrt{\frac{16\pi}{45}}Y_5^L.
\end{align}
\end{subequations}
Here the $Y_m^L$ are defined as functions of the spherical
harmonics of second order $Y^L_{2,n}$ \cite{zhou}:
\begin{subequations}
\begin{align}
Y_{1}^L & =\frac{i}{\sqrt{2}}\left(Y_{2,2}^L +
Y_{2,-2}^L\right),\nonumber
\\\nonumber
Y_{2}^L & =\frac{1}{\sqrt{2}}\left(Y_{2,-2}^L - Y_{2,2}^L\right) ,
 \\\nonumber
Y_{3}^L & =\frac{i}{\sqrt{2}}\left(Y_{2,1}^L+Y_{2,-1}^L \right),
\\\nonumber
Y_{4}^L & =\frac{1}{\sqrt{2}}\left(Y_{2,-1}^L-Y_{2,1}^L \right),
\\\nonumber
Y_{5}^L & = Y_{2,0}^L.
\end{align}
\end{subequations}

Equation (\ref{eq4}) can be written in terms of the real spherical
harmonics:
\begin{equation}\nonumber
R^L=
\frac{1}{2}\sqrt{\frac{16\pi}{15}}\left[Y_{1}^L\left(\frac{h_{11} -
h_{22}}{2}\right) + Y_{2}^Lh_{12} + Y_{3}^Lh_{13} + Y_{4}^Lh_{23} +
\frac{\sqrt{3}}{2}Y_{5}^L \right],
\end{equation}\nonumber
In  a more compact form this reads
\begin{equation}\label{harmonic}
R^L=\frac{1}{2}\sqrt{\frac{16\pi}{15}}\sum^5_1Y_{m}^Lh_m,
\end{equation}
where the $h_m$ are the quadrupolar representation  of the
gravitational wave:
\begin{equation}\nonumber
\begin{array}{cc}
h_1= & \left(\frac{h_{11} - h_{22}}{2}\right),
\\
h_2= & h_{12},
\\
h_3= & h_{13},
\\
h_4= & h_{23},
\\
h_5= & \frac{\sqrt{3}}{2}h_{33}.
\end{array}
\end{equation}
Therefore the matricial system of the equation (\ref{eq4}) takes
the form:
\begin{equation}\nonumber
\left[
\begin{array}{c}
R^1 \cr R^2 \cr R^3 \cr R^4 \cr R^5 \cr R^6
\end{array}
\right] =\frac{1}{2}\sqrt{\frac{16 \pi}{15}}
\begin{bmatrix}
Y_{1}^1 & Y_{2}^1 & Y_{3}^1 & Y_{4}^1 & Y_{5}^1 \cr Y_{1}^2 &
Y_{2}^2 & Y_{3}^2 & Y_{4}^2 & Y_{5}^2 \cr Y_{1}^3 & Y_{2}^3 &
Y_{2}^3 & Y_{4}^3 & Y_{5}^3 \cr Y_{1}^4 & Y_{2}^4 & Y_{3}^4 &
Y_{4}^4 & Y_{5}^4 \cr Y_{1}^5 & Y_{2}^5 & Y_{3}^5 & Y_{4}^5 &
Y_{5}^5 \cr Y_{1}^6 & Y_{2}^6 & Y_{3}^6 & Y_{4}^6 & Y_{5}^6 \cr
\end{bmatrix}
\left[
\begin{array}{c}
h_{1}\cr h_{2} \cr h_{3} \cr h_{4} \cr h_{5} \cr
\end{array}
\right],
\end{equation}
which is the same used as equation (\ref{solproinv3}).

\section{}\label{appen2}

We want to show in this appendix that equation (\ref{potential})
can be written in terms of the transducer's response. First we
notice that the potential of the tidal force density caused for
the passage of a gravitational wave is
\begin{equation}\nonumber
\phi({\bf x},t)=\sum_{j,k}\frac{1}{2}\rho x_j \ddot{h}_{jk}(t)x_k,
\end{equation}
where $\rho$ is the mass density and $x_i$ is the coordinate
location. The term $x_i$ represents a vector for the position so
it can be written of the following form:
\begin{equation}\nonumber
x_i = \epsilon n_i,
\end{equation}
where $\epsilon$ is the amplitude of the vector and $n_i$ is
represent the unit vector of the position vector. Replacing $x_i$
in the equation we get
\begin{equation}\nonumber
\phi({\bf x},t)=\sum_{j,k}\frac{1}{2}\epsilon^2 \rho n_j
\ddot{h}_{jk}(t)n_k,
\end{equation}
which can be written in the form
\begin{equation}\nonumber
\phi({\bf x},t)=\sum_{j,k}\frac{1}{2}\epsilon^2 \rho
\frac{\partial^2}{\partial t^2} \left(h_{jk}(t)n_jn_k\right).
\end{equation}

We have then the contraction of two tensors: the wave's tensor
$h_{ij}$ and a directionality tensor that characterizes the plan
in which the states of polarization of the gravitational wave act.
We can impose the traceless condition in $n_in_j$ adding the term
$-\frac{1}{3}\delta_{ij}$. Therefore the equation assumes the
form:
\begin{equation}\nonumber
\phi({\bf x},t)=\sum_{j,k}\frac{1}{2}\epsilon^2 \rho
\frac{\partial^2}{\partial t^2}
\left(h_{jk}(t)\left(n_jn_k-\frac{1}{3}\delta_{jk}\right)\right),
\end{equation}
where the term $\left(n_jn_k-\frac{1}{3}\delta_{jk}\right)$ is the
transducer's tensor of the equation (\ref{eq2}). Therefore we can
write
\begin{equation}\nonumber
\phi({\bf x},t)=\sum_{j,k}\frac{1}{2}\epsilon^2 \rho
\frac{\partial^2}{\partial t^2} \left(h_{jk}(t)D_{jk}\right),
\end{equation}
or in a  more summarized form:
\begin{equation}\nonumber \phi({\bf
x},t)=\frac{1}{2}\epsilon^2 \rho \ddot{\mathcal R},
\end{equation}
where $\mathcal R$ is the transducer's response of the equation
(\ref{eq1}). Using this equation we can write:
\begin{equation}\nonumber
\phi({\bf x},t)= \sqrt{\frac{4\pi}{15}}\epsilon^2 \rho \sum_m
\ddot{h}_m(t)Y_m.
\end{equation}

\end{document}